# Three-Dimensional Charge Density Wave Order in YBa$_2$Cu$_3$O$_{6.67}$ at High Magnetic Fields

S. Gerber[1,¶], H. Jang[2,¶], H. Nojiri[3], S. Matsuzawa[3], H. Yasumura[3], D. A. Bonn[4,5], R. Liang[4,5], W. N. Hardy[4,5], Z. Islam[6], A. Mehta[2], S. Song[7], M. Sikorski[7], D. Stefanescu[7], Y. Feng[7], S. A. Kivelson[8], T. P. Devereaux[1], Z.-X. Shen[1,8], C.-C. Kao[9], W.-S. Lee[1,*], D. Zhu[7,*], J.-S. Lee[2,*]

[1]*Stanford Institute for Materials and Energy Science, SLAC National Accelerator Laboratory and Stanford University, Menlo Park, California 94025, USA*

[2]*Stanford Synchrotron Radiation Lightsource, SLAC National Accelerator Laboratory, Menlo Park, California 94025, USA*

[3]*Institute for Materials Research, Tohoku University, Katahira 2-1-1, Sendai, 980-8577, Japan*

[4]*Department of Physics & Astronomy, University of British Columbia, Vancouver, British Columbia, Canada V6T 1Z1*

[5]*Canadian Institute for Advanced Research, Toronto, Ontario, Canada M5G 1Z8*

[6]*The Advanced Photon Source, Argonne National Laboratory, Argonne, Illinois 60439, USA*

[7]*Linac Coherent Light Source, SLAC National Accelerator Laboratory, Menlo Park, California 94025, USA*

[8]*Geballe Laboratory for Advanced Materials, Departments of Physics and Applied Physics, Stanford University, Stanford, California 94305, USA*

[9]*SLAC National Accelerator Laboratory, Menlo Park, California 94025, USA*

[¶]These authors contributed equally to this work.

[*]Correspondence to: dlzhu@slac.stanford.edu, leews@stanford.edu, jslee@slac.stanford.edu

**Charge density wave (CDW) correlations have recently been shown to universally exist in cuprate superconductors. However, their nature at high fields inferred from nuclear magnetic resonance is distinct from that measured by x-ray scattering at zero and low fields. Here we combine a pulsed magnet with an x-ray free electron laser to characterize the CDW in $YBa_2Cu_3O_{6.67}$ via x-ray scattering in fields up to 28 Tesla. While the zero-field CDW order, which develops below $T \sim 150$ K, is essentially two-dimensional, at lower temperature and beyond 15 Tesla, another three-dimensionally ordered CDW emerges. The field-induced CDW onsets around the zero-field superconducting transition temperature, yet the incommensurate in-plane ordering vector is field-independent. This implies that the two forms of CDW and high-temperature superconductivity are intimately linked.**

The universal existence of charge density wave (CDW) correlations in superconducting cuprates (*1-7*) raises profound questions regarding the role of charge order – is it competing or more intimately intertwined with high-temperature superconductivity (HTSC) (*8-10*)? Uncovering the evolution of CDW order upon suppression of HTSC by an external magnetic field provides valuable insight into this question. One of the most studied cuprate superconductors, $YBa_2Cu_3O_{6+\delta}$, has become a model material for the study of CDW phenomena in cuprates. Incommensurate CDW order has recently been found to coexist with HTSC using x-ray scattering measurements (*2, 3, 11, 12*). The temperature and magnetic field dependencies up to $\mu_0H = 17$ T indicate a competition between CDW order and HTSC (*3, 11*). However, nuclear magnetic resonance (NMR) (*13*) and Hall coefficient (*14*) measurements at even higher magnetic fields imply a much lower CDW onset temperature than those found via x-ray scattering measurements (*12*). These discrepancies have led to speculations on the emergence of different CDW correlations at high magnetic fields (*15*), which are also supported by ultrasonic (*16*) and torque magnetometry measurements (*17*). If so, it is crucial to reveal the structure of the CDW at high fields and to track how it emerges in the presence of the zero-field CDW. However, so far the highest magnetic field strengths available for x-ray scattering techniques are not sufficient to detect CDW order in superconducting cuprates beyond 17 T.

To gain insight into this critical question one needs to bridge this technological gap by innovating the experimental approach. Here, we employ x-ray scattering in the presence of pulsed high magnetic fields using an x-ray free electron laser (FEL). The unprecedented brilliance of the Linac Coherent Light Source (LCLS) (*18*) enables the measurement of the weak CDW scattering signal with a single x-ray pulse (~ 50 femtosecond) at the apex of a millisecond magnetic field pulse (*19*). This approach provides us with the opportunity to probe the CDW signal in YBa$_2$Cu$_3$O$_{6+\delta}$ at magnetic fields beyond 17 T, thereby entering a field range comparable to that used in NMR (*13*), Hall coefficient (*14*) and ultrasonic measurements (*16*).

Figure 1A shows a schematic of how the two pulsed sources – the magnet and the x-ray FEL – were synchronized to study the CDW in detwinned, underdoped YBa$_2$Cu$_3$O$_{6.67}$ (YBCO) with ortho-VIII oxygen order (*19*). To monitor the field dependence of the CDW, an area detector was used to capture a cut of the *kl*-plane in reciprocal space. The full view of the zero-field diffraction pattern in the vicinity of the CDW *Q*-vector position at the zero-field superconducting transition temperature, $T_c(H = 0) = 67$ K, is shown in Fig. 1B. In this geometry, we observe CDW features centered around ***Q*** = (0, 2-*q*, ±½) with an incommensuration $q \sim 0.318$ (*2, 3, 11, 12*). The detected diffraction pattern of the CDW shows that the correlation along the crystallographic *c*-axis is very weak, resulting in a rod-like shape along the *l*-direction. Moreover, we also measured the temperature dependence of the zero-field CDW (Fig. 1C and Ref. [*19*]), reproducing earlier reports that the CDW signal is maximal at $T_c$ and suppressed for $T < T_c$ (*2, 3, 11, 12*), which indicates a competition between CDW order and HTSC.

We first discuss the temperature dependence of the CDW at $\mu_0 H = 20$ T. Figure 2A shows the (0, 2-*q*, *l*) CDW signal at both 0 and 20 T. There is no field-induced change of the CDW at $T_c$, which is consistent with earlier results (*3*). With decreasing temperature ($T < T_c$), the CDW signal becomes sharper along the *k*-direction and more intense than at zero field. This indicates that, as the magnetic field suppresses superconductivity, the CDW order is enhanced (Fig. 2B). Surprisingly, as shown in the 2D difference map $I_{20T} - I_{0T}$ (lower panels of Fig. 2A) the field-induced enhancement is most dramatic at $l \sim 1$, rather than at $l \sim ½$ where the zero-field CDW signal is maximal (*2, 3, 11, 12*). This observation indicates that a new kind of CDW correlation emerges around $T_c(0)$ − well below the zero-field CDW onset temperature (Fig. 1C). As shown in Fig. 2C, the temperature dependence of the field-induced CDW is consistent with that of the CDW signatures inferred from NMR measurements (*13*), implying that both share the same origin.

Next we explore the field-induced enhancement of CDW order at $T = 10$ K. Figure 3A shows the diffraction patterns (upper panels) at $\mu_0 H = 0 - 25$ T. The lower panels depict the projected intensities at both $l \sim \frac{1}{2}$ and $l \sim 1$, *i.e.*, integrated over the indicated range of $l$. Up to $\mu_0 H = 15$ T the intensities of the CDW order at both $l \sim \frac{1}{2}$ and $l \sim 1$ are similar. Above 15 T, however, the intensity at $(0, 2-q, \sim 1)$ continues to grow strongly, while it saturates at $(0, 2-q, \sim \frac{1}{2})$ (Fig. 3B). This was confirmed at an equivalent CDW $Q$-vector $(0, 2+q, 1)$ (Fig. 4 and Ref. [*19*]), where we were able to follow the enhancement of CDW intensity at $l \sim 1$ up to our maximum field, $\mu_0 H = 28$ T. Furthermore, the in-plane correlation lengths $\xi_k$ at both $l \sim \frac{1}{2}$ and $l \sim 1$ also differentiate as a function of $H$ (Fig. 3C), suggestive of a transition; $\xi_k$ at $l \sim 1$ increases for $\mu_0 H > 15$ T, while $\xi_k$ at $l \sim \frac{1}{2}$ saturates or is slightly suppressed. We note that the estimated correlation length at the highest magnetic fields may be limited by the instrument resolution. Nevertheless, the distinct field-dependence of the CDW intensity and the correlation length confirm that the CDW order at $l \sim 1$ is different from that at $l \sim \frac{1}{2}$, and that both CDW orders coexist at high magnetic fields. In particular, the onset of the field-induced CDW ($l \sim 1$) at $\mu_0 H > 15$ T is consistent with NMR results in which the line-splitting signature of CDW order is absent at low fields (*13*).

Data shown in Fig. 3 motivates scrutiny of the field-induced CDW in the $l \sim 1$ region at the highest accessible magnetic field 28 T. Given our experimental configuration (*19*), a larger $l$-range is accessible near $l = 1$ by monitoring the equivalent CDW reflection near $(0, 2+q, l)$, rather than near $(0, 2-q, \sim 1)$. As shown in Figs. 4A and 4B, the CDW diffraction pattern at 28 T becomes sharper not only along the $k$-direction (Fig. 3C), but also along the $l$-direction (perpendicular to the $CuO_2$ planes). This indicates that CDW correlations along the $c$-axis are enhanced, i.e. $\xi_l = 50(2)$ Å at 28 T, concomitant with roughly a three-fold increase of the peak height. Therefore, the field-induced CDW at $l = 1$ is much more correlated in all three dimensions than the zero-field CDW. Importantly, as shown in Figs. 4C and 4D, the CDW peak positions are identical at 20 and 28 T. There has been speculation that the in-plane component of the CDW $Q$-vector may shift and lock-in to a commensurate value at high magnetic fields (*20*). However, within our experimental resolution the field-induced $h$- and $k$-components of the $Q$-vector are identical to that of the zero-field CDW.

We note that a field-induced spin density wave (SDW) has been observed in $La_{2-x}Sr_xCuO_4$ at weaker fields $\sim 6$ T, which is also peaked at integer $l$ due to an alignment of SDW patches, associated with the vortex cores (*21*). However, the emergence of field-induced CDW order at $l = 1$ is unlikely due to the alignment of CDW regions that are associated with vortices (*22*). This is because at magnetic fields beyond 20 T, the distance between vortices, if still existent, would be less than ~100 Å in the $CuO_2$ plane (*23*), which is already smaller than the in-plane CDW correlation length at these field strengths (Fig. 3C).

Now we discuss implications of the observed field-induced three-dimensional CDW at $l = 1$. First, its emergence at high fields and low temperatures implies a boundary that separates the phase diagram into different CDW regions, as also suggested by ultrasonic (*16*) and NMR measurements (*24*). Second, given that a field-dependence of the CDW order is only observed below $T_c(0)$ (Fig. 2), we infer that the enhancement is related to the suppression of superconductivity. Thus, the growth of the CDW peak intensity up to 28 T indicates that superconducting correlations still exists beyond the upper critical field that was deduced from transport measurements (*25, 26*). Third, our observations shed new light on quantum oscillation results, which have been interpreted as evidence for the existence of small electron pockets in the "nodal" region of the Brillouin zone (*27-29*). It is plausible that the Fermi surface is reconstructed by the stronger field-induced CDW at $l = 1$, rather than the shorter-range correlated one at $l \sim ½$ (*30*). Finally, we remark that the relation between the zero-field and field-induced CDW is puzzling. On the one hand, they seem unrelated as they exhibit distinct temperature and field dependences, as well as a different ordering perpendicular to the $CuO_2$ plane. On the other hand they must be somehow related, since they feature the same in-plane CDW incommensuration $q$. Thus, our results reveal a rich CDW phenomenonology in cuprates, which is not a simple competition with HTSC.

**Acknowledgments**

We thank J. Hastings, J. Defever, D. Damiani, G. Curie and V. Borzenet for technical assistance in developing the scattering setup. Discussions with C. Mielke are acknowledged. This work was supported by the Department of Energy (DOE), Office of Science, Basic Energy Sciences, Materials Sciences and Engineering Division, under contract DE-AC02-76SF00515. X-ray FEL studies were carried out at the Linac Coherent Light Source, a Directorate of SLAC and an Office of Science User Facility operated for the U.S. DOE, Office of Science by Stanford University. Soft and hard x-ray scattering studies were carried out at the Stanford Synchrotron Radiation Lightsource, a Directorate of SLAC and an Office of Science User Facility operated for the U.S. DOE, Office of Science by Stanford University. Hard x-ray scattering studies were also conducted at the Advanced Photon Source, supported by the U.S. DOE, Office of Science, Office of Basic Energy Sciences under Contract No. DE-AC02-06CH11357. S.G. acknowledges partial support by the Swiss National Science Foundation under Fellowship No. P2EZP2_148737. H.N. acknowledges the support by KAKENHI 23224009, ICC-IMR and MD-program. Materials development was supported by the Natural Sciences and Engineering Research Council, and by the Canadian Institute for Advanced Research.

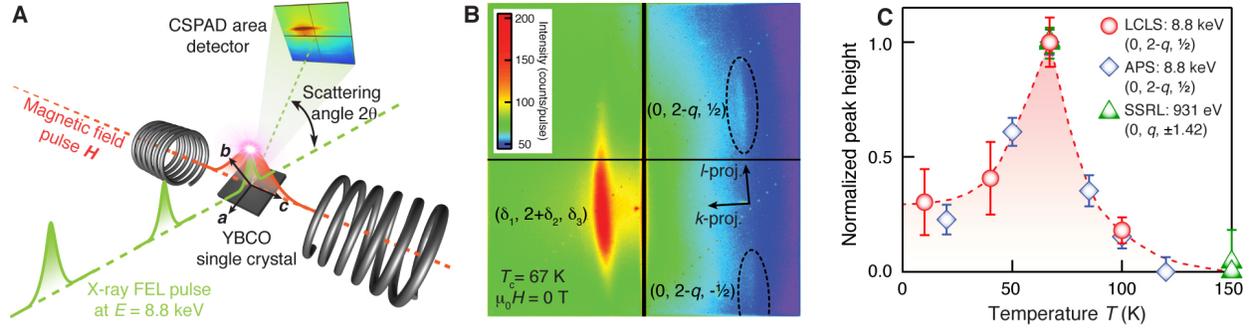

**Fig. 1. Experimental setup and zero-field characterization.** (**A**) The millisecond pulsed magnetic field and femtosecond x-ray FEL pulses are synchronized to obtain a diffraction pattern from the YBCO single crystal at the maximum magnetic field. The diffraction pattern was recorded using a 2D pixel array detector. (**B**) Zero-field diffraction pattern showing the (0, 2-$q$, ±½) CDW peaks and the tail of the (0, 2, 0) Bragg peak ($\delta_1$= -0.118, $\delta_2$ = 0.001, $\delta_3$ = 0.021). The sample rotation angle was optimized for the CDW position and not for the (0, 2, 0) Bragg peak (*19*). (**C**) The temperature dependence of the CDW peak height near (0, 2-$q$, ½) is shown with red colored symbols (LCLS). Data measured at synchrotron lightsources using both non-resonant (APS) and resonant (SSRL) x-ray scattering are also shown for comparison. The dashed line is a guide-to-the-eye and the error bars denote 1 standard deviation (s.d.) as obtained from the peak fitting.

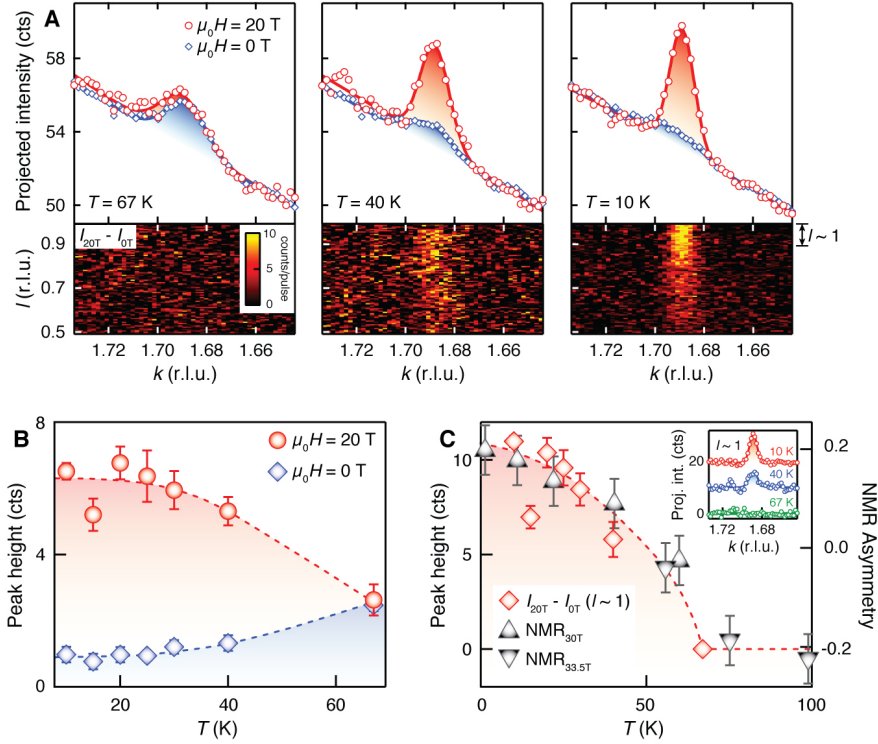

**Fig. 2. Temperature dependence of the CDW order at $\mu_0 H = 20$ T.** (**A**) The upper and lower panels show the evolution of the projected $(0, 2\text{-}q, \frac{1}{2})$ CDW peak profile along the $k$-direction and the difference map of the diffraction pattern between $\mu_0 H = 0$ and 20 T, respectively, at representative temperatures of $T = 67$, 40, and 10 K. Positions are given in reciprocal lattice units (r.l.u.). Solid lines are Gaussian fits to the data with a 2$^{nd}$ order polynomial background. (**B**) $T$-dependence of the peak height from the projected CDW profiles at 0 and 20 T. (**C**) Peak height of the projected CDW profiles near $l \sim 1$ as a function of temperature. The projected CDW profiles (inset, traces offset by 10 cts) are obtained from the 2D difference map by integrating near $l \sim 1$, as indicated in (**A**). As a comparison NMR data (*13*) are superimposed. Dashed lines are guides-to-the-eye. Error bars correspond to 1 s.d.

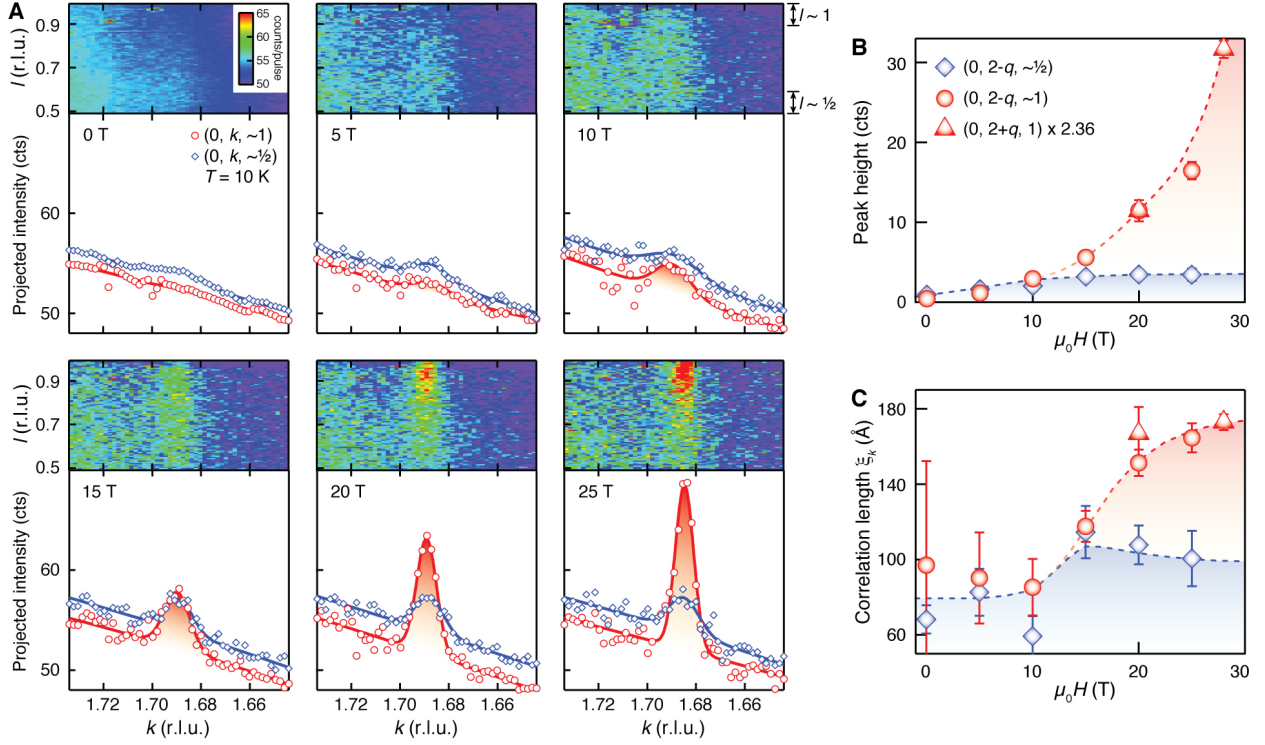

**Fig. 3. Field dependence of the CDW order at $T$ = 10 K.** (**A**) CDW diffraction pattern (upper panels) and projected CDW peak profiles (lower panels) near $l \sim \frac{1}{2}$ and $l \sim 1$, obtained by integration of the signal in the windows indicated on the image, in the field range $\mu_0 H = 0 - 25$ T. Features due to ice condensation on the sample surface, which do not overlap with the CDW signal, were subtracted from the diffraction patterns (*19*). Solid lines are Gaussian fits to the data with a 2$^{nd}$ order polynomial background. (**B**) Peak height of the projected CDW profile near $l \sim \frac{1}{2}$ and $l \sim 1$ as a function of $H$. Data taken at an equivalent CDW region $(0, 2+q, 1)$, shown in Fig. 4, are superimposed by normalizing the values at 20 T. (**C**) $H$-dependence of the in-plane correlation length $\xi_k = 1/\sigma_k$ deduced from Gaussian fits ($\sigma_k$ is the Gaussian standard deviation) to the projected CDW profile shown in (**A**) as well as Fig. 4C. Values of $\xi_k$ have not been corrected for the instrument resolution and, therefore, represent lower bounds. Dashed lines are guides-to-the-eye. Error bars correspond to 1 s.d.

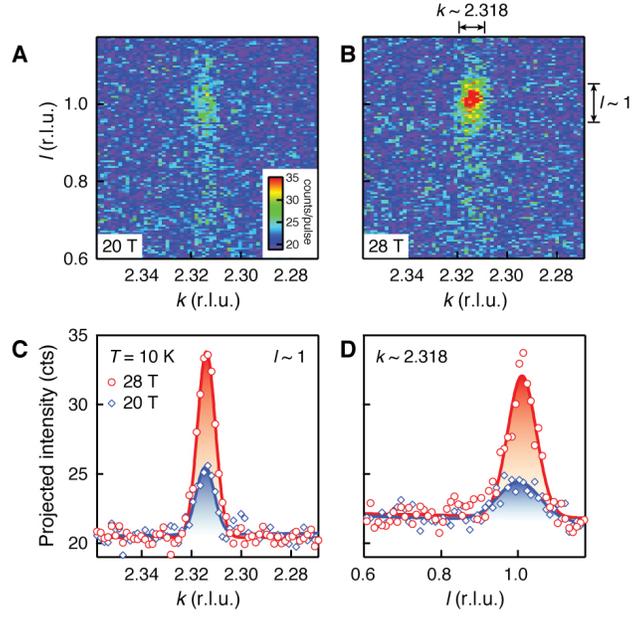

**Fig. 4. Three-dimensional CDW order at μ₀H > 20 T.** (**A** and **B**) CDW diffraction pattern near (0, 2+q, l) at $\mu_0 H$ = 20 and 28 T. (**C** and **D**) Projected CDW peak profiles along the *k*- and *l*-direction within the regions indicated in (**B**). Gaussian fits to the data with a linear background (solid lines), reveal that the field-induced CDW peak is centered at *k* = 2.318(1) and *l* = 1.00(1).

# Supplementary information

## Sample preparation and characterization

A detwinned $YBa_2Cu_3O_{6.67}$ single crystal with ortho-VIII oxygen order and a zero-field superconducting transitions temperature of $T_c = 67$ K was used for this study. The dimensions of the single crystal were 1.1 mm × 0.5 mm × 0.6 mm along the crystallographic *a*, *b,* and *c*-axis, respectively (lattice parameters: $a = 3.82$ Å, $b = 3.88$ Å, and $c = 11.68$ Å). We prepared a thin crystal along the *b*-axis to limit sample heating via eddy currents due the pulsed magnetic field. The scattering surface was normal to the *b*-axis, allowing to measure CDW peaks close to the (0, 2, 0) structural Bragg peak (Fig. S2A), whilst applying the magnetic field *H* along the *c*-axis.

The single crystal was thoroughly characterized by measuring the zero-field CDW signal at the Advanced Photon Source (APS), beamline 6-ID-B and the Stanford Synchrotron Radiation Lightsource (SSRL), beamlines 7-2, 10-2, and 13-3, using both hard x-ray scattering measurements at $E = 8.8$ keV and resonant soft x-ray scattering measurements at the Cu $L_3$-edge, $E = 931$ eV (Figs. 1C and S1).

## Experimental setup at LCLS

The x-ray FEL experiment described in the main text was performed at the X-ray Correlation Spectroscopy (XCS) instrument of the Linac Coherent Light Source (LCLS) at the SLAC National Accelerator Laboratory in its pink beam configuration (*31*). A photon energy of $E = 8.8$ keV (sigma/horizontally polarized x-rays) was chosen, specifically below the Cu *K*-edge to reduce fluorescence background. Beryllium compound refractive lenses were used to focus the beam to match the size of the crystal. Accounting for the beamline transmission losses, a single x-ray pulse flux was estimated to be $\sim 1 \times 10^{11}$ photons on the sample. A 2D pixel array CS-140k detector (*32*) was used to record the scattered x-rays.

A split-pair pulsed magnet with an inner bore size of 3 mm, 20 mm outer diameter and 1.4 mm wide exit windows was used to generate the magnetic field pulse (*33*). The YBCO single crystal was mounted at the end of a 1.5 mm diameter sapphire crystal rod and positioned at the center of the magnet (Fig. S2B). The *c*-axis of the sample was aligned along the magnetic field direction with a tilt of ~4º, allowing scattering from the CDW through the 1.4 mm magnet window (Figs. S2C and S2D). Accessible scattering vectors in reciprocal space were estimated by the scattering geometry and the detector position.

As shown in Fig. S3, the magnetic field pulse duration was ~1 ms with a 0.5 ms rise time. The ~50 fs duration x-ray pulse was synchronized via a gate trigger with the magnetic field pulse to measure the diffraction pattern at the maximum field strength. The highest pulsed field, $\mu_0 H = 28$ T, was generated by a charging voltage of 980 V and peak current of 2860 A. Ohmic heating of the magnet coils led to waiting times of 2 to 25 minutes between subsequent pulses for field strengths from $\mu_0 H = 5$ to 28 T. The coils were cooled by a displex closed-cycle cryostat. The sample assembly was mounted on a separated displex, allowing for independent sample motion as well as temperature control ($T = 10 - 400$ K).

Figure S4 shows the stability of the experimental setup under pulsed magnetic fields. The position of the (0, 2, 1) structural Bragg peak at $\mu_0 H = 0$ and 20 T remains stable within the experimental error, which is well below the width of the CDW peak. Moreover, it is important to stress that the estimated correlation lengths have not been corrected for the instrument resolution and, therefore, represent lower bounds. Under our experimental conditions the instrument resolution was dominated by the bandwidth due to the jitter of the x-ray FEL and the actual width of the diffraction peak.

During measurements a vacuum of ~$10^{-6}$ Torr was obtained, which upon cooling resulted in condensation of water molecules on the cryostat cold finger and the sample surface. The crystalized ice appears in form of a ring in the diffraction pattern at $2\theta \sim 38°$ for $E = 8.8$ keV (*34*). Accidentally, this ice ring is located in the vicinity of the (0, 2-*q*, *l*) CDW position ($2\theta \sim 35.8°$), but does not overlap with the CDW signal in the probed *l*-range. Some of the raw data, depending on the experimental conditions such as temperature cycling and magnet operation, featured such an "ice ring" (Fig. S5A). We confirmed that warming the sample eliminates the ice ring and yields a pristine diffraction pattern, such as the one depicted in Fig. 1B. To extract the projected CDW intensity as shown in Figs. 2A and 3A, the ice ring was fitted with a Gaussian curve for each *l*-value and subtracted from the raw data (Fig. S5B).

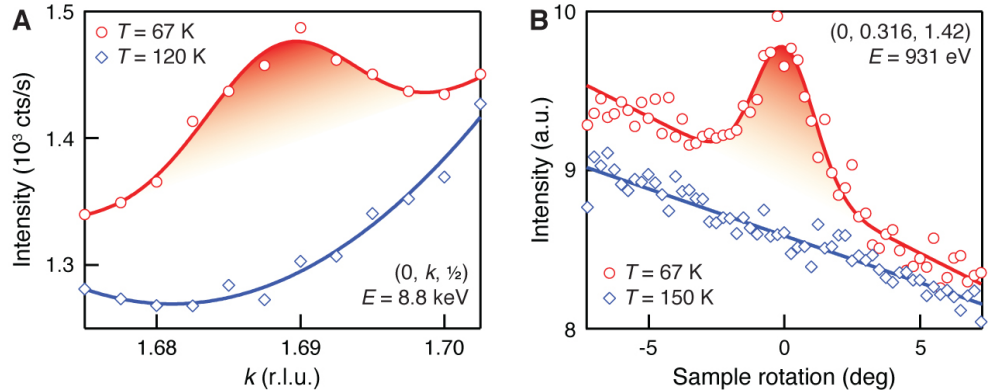

**Figure S1:** (**A**) Zero-field CDW peak measured using $E = 8.8$ keV hard x-ray scattering at APS. The signal is strongest at $T_c = 67$ K (red) and absent at $T = 120$ K (blue, offset by -100 cts/s). (**B**) Rocking curve of the (0, 0.316, 1.42) zero-field CDW peak, measured at the Cu $L_3$-edge ($E = 931$ eV) using resonant soft x-ray scattering at SSRL. An $l$-value of 1.42 r.l.u., i.e., slightly less than the half-integer value, had to be chosen due to experimental constraints and the limited total momentum transfer at $E = 931$ eV. Lines are Gaussian fits to the data with a 2$^{nd}$ order polynomial (APS) and linear (SSRL) background.

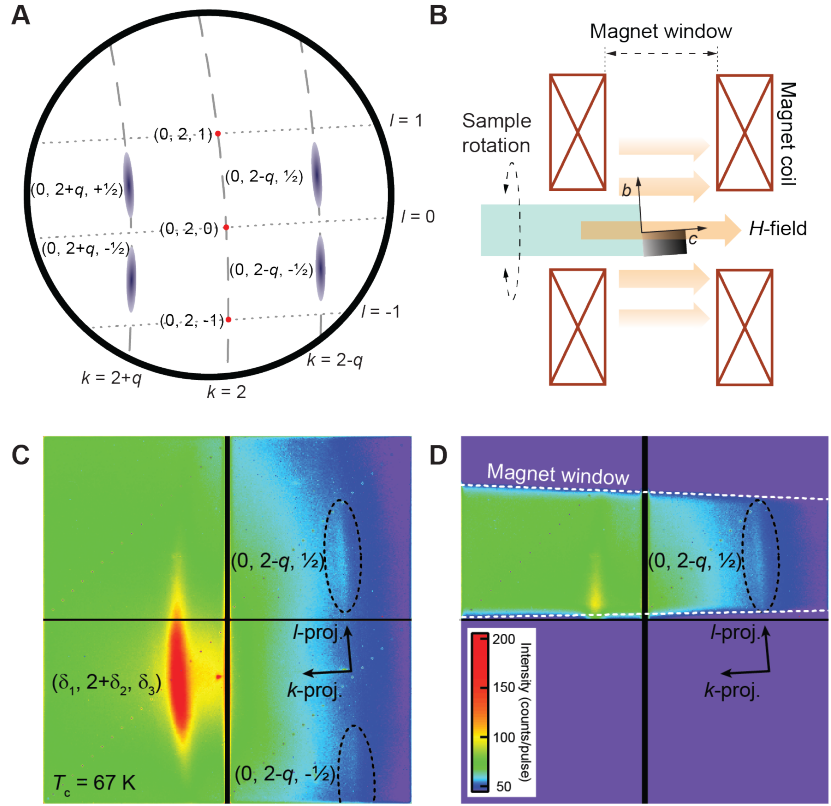

**Figure S2:** (**A**) Accessible (0, $k$, $l$) reciprocal space in our scattering geometry (black circle). Red dots and blue ellipses indicate the positions of structural and CDW Bragg peaks, respectively. The structural Bragg peaks, (0, 2, 0) and (0, 2, ±1), were used as references to explore the CDW signal. (**B**) Experimental geometry of the pulsed magnet and the sample with the sample rotation axis parallel to $H$. The $b$-axis of the YBCO single crystal was tilted by ~4º with respect to the magnetic field to fit the (0, 2±$q$, $l$) CDW signal within the magnet window. (**C**) Zero-field diffraction pattern without magnet showing the (0, 2-$q$, ±½) CDW peaks and the tail of the (0, 2, 0) Bragg peak ($\delta_1$ = -0.118, $\delta_2$ = 0.001, $\delta_3$ = 0.021). (**D**) In the presence of the magnet, scattered x-rays are blocked by the split-pair coils except for the region near the (0, 2-$q$, ½) CDW peak, which is detectable through the magnet exit windows (measured at $T$ = 67 K and $\mu_0 H$ = 0 T).

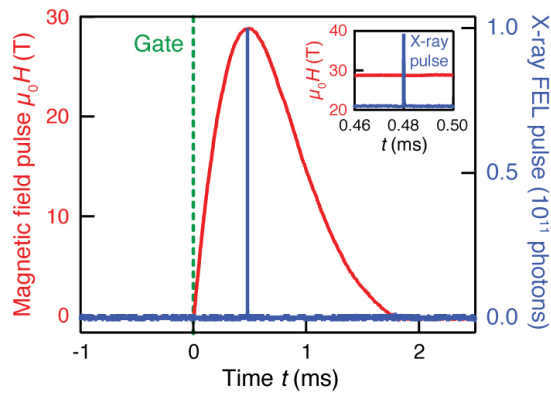

**Figure S3:** Time trace of the millisecond magnetic field pulse (red line) and the femtosecond x-ray pulse (blue line). Synchronization of the two pulses via a gate trigger (green dotted line) permits measurements at the maximum field of $\mu_0 H = 28$ T with $\sim 10^{11}$ photons from the x-ray FEL (inset).

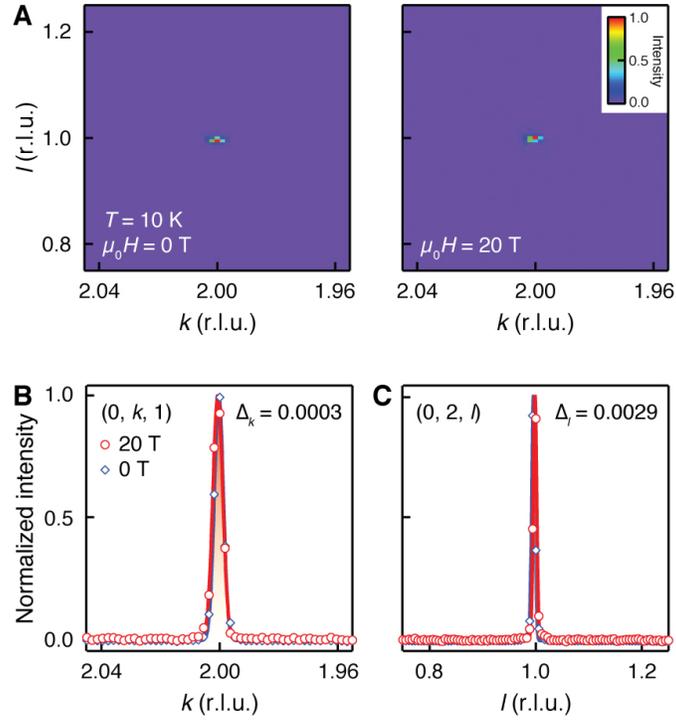

**Figure S4:** (**A**) Diffraction pattern of the (0, 2, 1) structural Bragg peak at $\mu_0 H = 0$ and 20 T. (**B**) and (**C**) show the normalized projected Bragg peak intensity along the $k$- and $l$-direction, respectively. Gaussian fits to the projected intensities reveal that the (0, 2, 1) structural Bragg peak moves only by $\Delta_k = 0.0003$ and $\Delta_l = 0.0029$ r.l.u. in the presence of a 20 T magnetic field.

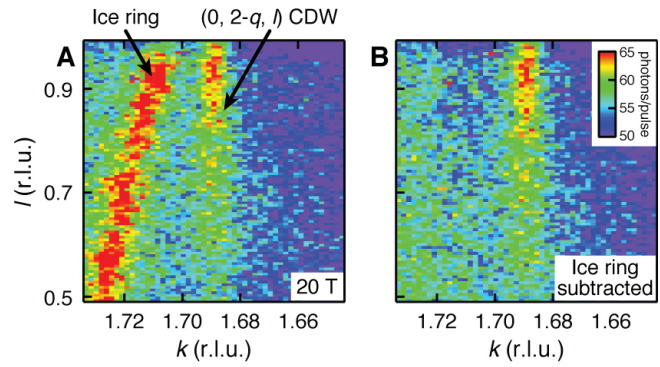

**Figure S5:** (**A**) Raw diffraction data measured at $\mu_0 H = 20$ T and $T = 10$ K near the $(0, 2\text{-}q, l)$ CDW signal. Condensation of water molecules results in the formation of an "ice ring" in the vicinity of the CDW signal. (**B**) Fitting of the ice ring by a Gaussian curve for each $l$-value and subtracting it from the raw data, allows to recover the pristine diffraction pattern.